\documentclass[twocolumn]{autart}    

\usepackage{graphicx}          

\usepackage{mathtools} 
\usepackage{amssymb,amsfonts}
\usepackage{bm}        
\usepackage{mathrsfs}  
\usepackage[normalem]{ulem}
\newtheorem{assu}{Assumption}
\newtheorem{lemma}{Lemma}

\begin{document}

\begin{frontmatter}

\title{Distributed Omniscient Observers for Multi-Agent Systems: Design and Applications\thanksref{footnoteinfo}} 

\thanks[footnoteinfo]{This research is supported by the National Research Foundation, Singapore, and PUB, Singapore’s National Water Agency under its RIE2025 Urban Solutions and Sustainability (USS) (Water) Centre of Excellence (CoE) Programme, awarded to Nanyang Environment \& Water Research Institute (NEWRI), Nanyang Technological University (NTU), Singapore. This research is also supported by the Ministry of Education, Singapore, under its Academic Research Fund Tier 1 (RG95/24). Any opinions, findings and conclusions or recommendations expressed in this material are those of the author(s) and do not reflect the views of the National Research Foundation, Singapore and PUB, Singapore’s National Water Agency. The material in this paper was not presented at any conference. Corresponding author: Xunyuan Yin.}

\author[NEWRI]{Ganghui Cao}\ead{ganghui.cao@ntu.edu.sg},   
\author[NEWRI,CCEB]{Xunyuan Yin}\ead{xunyuan.yin@ntu.edu.sg}

\address[NEWRI]{Nanyang Environment \& Water Research Institute, Nanyang Technological University, 1 CleanTech Loop, 637141, Singapore}  
\address[CCEB]{School of Chemistry, Chemical Engineering and Biotechnology, Nanyang Technological University, 62 Nanyang Drive, 637459, Singapore}     
          

\begin{abstract}                          
This paper proposes distributed omniscient observers for both heterogeneous and homogeneous linear multi-agent systems, such that each agent can correctly estimate the states of all agents. The observer design is based on local input-output information available to each agent, and knowledge of the global communication graph among agents is not necessarily required. The proposed observers can contribute to distributed Nash equilibrium seeking in multi-player games and the emergence of self-organized social behaviors in artificial swarms. Simulation results demonstrate that artificial swarms can emulate animal social behaviors, including sheepdog herding and honeybee dance-based navigation.
\end{abstract}

\end{frontmatter}

\section{Introduction}

Consensus is widely recognized as one of the most fundamental cooperative behaviors in multi-agent systems (MAS). Analogous to those biological synchronies (e.g., synchronous flashing of fireflies \cite{John1968}), it typically describes a phenomenon that the state trajectories of all agents evolve identically. 

Since the pioneering research in \cite{Fax2002,Jadbabaie2003,Saber2003}, consensus control of MAS has been extensively investigated over the past two decades. 
Most research formulates the consensus problem within a distributed context, where each agent only has access to limited local information and can only communicate with neighboring agents \cite{W.Ren2008book,Z.Li2017book}.
Numerous consensus protocols have been developed under various complications, including model uncertainties \cite{Zhongkui2014,Junjie2016}, switching and disturbed communication links \cite{Wei2005,Zhongkui2017,Guanghui2019}, velocity and acceleration constraints \cite{Peng2017,Junjie2018,Junjie2019,Junjie2020}, etc. 

As consensus-reaching has been relatively well studied, recent focus has been increasingly placed on more advanced forms of cooperation in MAS. However, the aforementioned distributed setup poses challenges to achieving advanced collaboration. Specifically, if an agent only has access to limited local information, it may fail to effectively cooperate with others for a global objective of the MAS. This motivates the development of distributed omniscient observers in this paper, which aim to provide each agent with sufficient global information to support autonomous decision-making.

In prior research on leader–follower MAS, distributed observers have been commonly designed either for each follower to estimate the leader’s state \cite{Yiguang2008,Jie2017,Tao2019,S.Wang2021,S.Wang2023,T.Liu2024,Jixing2026}, or for each agent to reconstruct its own absolute state using relative output information \cite{H.Zhang2011}. Different from the above approaches, the distributed observers developed in \cite{T.Kim2020} allow each vehicle in a formation to reconstruct the attitudes and positions of all vehicles. This establishes a prototype for distributed global-state estimation in MAS. However, this result applies only to the zero-input case, which may limit its applicability in practical scenarios where agents are under active control or subject to external inputs.

In contrast, the distributed omniscient observers proposed in this paper enable every agent to estimate the global state of MAS, while allowing each agent’s input to be persistently nonzero.
This equips each agent with a ``global view'', enabling the MAS to cooperatively perform
complex tasks that extend beyond basic consensus.
The proposed observer design method mainly builds on the framework developed in \cite{Ganghui2025arXiv}. However, the method in this paper emphasizes exploiting relative (neighbor-to-neighbor) output information in MAS, which is often more reliable and easier to realize in practice compared with using the absolute one\footnote{For example, in relative localization problems, only a small number of agents obtain their absolute positions from the Global Navigation Satellite System (GNSS), while the remaining agents measure relative positions with respect to their neighbors. This setup offers advantages under GNSS non-line-of-sight conditions and in GNSS-denied or jammed environments \cite{Zhimin2019,XiaoShen2024}.}.
More detailed technical advancements over \cite{Ganghui2025arXiv} are clarified at the beginning of Section~~\ref{DOOD}.

\subsection*{Notation}
For a vector $x$, $\left\| x \right\|$ denotes the Euclidean norm.
For a matrix $X$, ${\rm{Im}}X$ denotes the range or image of $X$. 
${\rm{Re}}\lambda (X) < 0$ indicates that all eigenvalues of $X$ lie in the open left half of the complex plane.
If $X = X^\top$, $\lambda_{\min} (X)$ denotes the smallest eigenvalue of $X$.
$I$ and $0$ denote the identity matrix and zero matrix of appropriate dimensions, respectively.
For a collection of matrices $\left\{X_i|\ i=1,2,\cdots,N\right\}$, ${\rm{diag}}({X_i})_{i = 1}^N$ denotes the block-diagonal matrix formed by $X_i$, and ${\rm{col}}{({X_i})_{i = 1}^N}$ denotes the matrix obtained by stacking them, i.e., 
$\setlength\arraycolsep{1.4pt}
{\left[ {\begin{array}{*{20}{c}}
			{X_1^{\top}}&{X_2^{\top}}& \cdots &{X_N^{\top}}
	\end{array}} \right]^{\top}}$, provided that the dimensions are compatible.

\section{Review of the Distributed Observers Without Using Global Inputs}\label{RDOWUGI}

\subsection{Communication Graph}
The communication links among observer nodes in distributed observers (or among agents in MAS) enable information exchange.
The topology of the communication links can be characterized using an undirected graph, introduced as follows.

A graph $\mathcal{G}=(\mathcal{N},\mathcal{E},\mathcal{A})$ consists of a finite, nonempty node set $\mathcal{N}=\left\{1,2,\cdots,N\right\}$, an edge set $\mathcal{E}\subseteq \mathcal{N}\times \mathcal{N}$ whose elements are ordered pairs of nodes, and an adjacency matrix $\mathcal{A}=[a_{ij}]\in \mathbb{R}^{N\times N}$. An edge originating from node $j$ and ending at node $i$ is denoted by $(j, i)\in \mathcal{E}$, which represents a directed information flow from node~$j$ to node~$i$. 
The adjacency matrix is formed by the weights of edges, with $a_{ij}>0$ if $(j,i)\in \mathcal{E}$, and $a_{ij}=0$ otherwise.
We assume that the graph has no self-loops, i.e., $a_{ii}=0$, $\forall i \in \mathcal{N}$. The Laplacian matrix $\mathcal{L}=[l_{ij}]\in \mathbb{R}^{N\times N}$ of graph $\mathcal{G}$ is defined by $l_{ii}=\sum\nolimits_{k = 1}^N {{a_{ik}}}$ and $l_{ij}=-a_{ij},\ \forall i,j \in \mathcal{N},\ i\neq j$. 
A directed path from node $i$ to node $j$ is a sequence of edges $(i_{k-1},\ i_{k})\in \mathcal{E},\ k=1,2,\cdots,\bar k$, where $i_0=i,\ i_{\bar k}=j$. 
Graph $\mathcal{G}$ is said to be undirected if $a_{ij} = a_{ji}, \forall i,j \in \mathcal{N}$. An undirected graph is said to be connected if, for every pair of distinct nodes $i,j\in\mathcal{N},\ i\neq j$, there exists at least one directed path from node $i$ to node $j$.

\subsection{Revisiting the Design Method}\label{DMR}

In \cite{Ganghui2025arXiv}, a distributed observer design method was proposed for the following linear time-invariant system:
\begin{equation}
    \dot x = Ax + Bu, \label{LTIsys}
\end{equation}
where $x\in \mathbb{R}^{n}$, $u\in \mathbb{R}^{m}$ are the system state and input, respectively.
The distributed observers comprise $N$ observer nodes, and each node aims to estimate the global system state $x$. For the $i$th observer node, it has access to a local output
\begin{equation}
    y_i = C_i x,  \label{yidef}
\end{equation}
where $y_i\in \mathbb{R}^{p_i}$, and $C_i$ has full row rank.
In addition, it accesses a local input $u_i\in \mathbb{R}^{m_i}$, which only contains partial information of the global input $u$. Specifically,
\begin{equation}\label{uiconstraint}
    Bu = B_i u_{i} + B_{-i} u_{-i},
\end{equation}
where $u_{-i}\in \mathbb{R}^{m_{-i}}$ denotes the input information unavailable to the $i$th node.
From $A$, $C_i$, and $B_{-i}$, the following four matrices can be computed:
\begin{equation*}
    {T_{id}} \in \mathbb{R}^{n \times \delta_i },\ {E_i} \in {\mathbb{R}^{\delta_i  \times \delta_i }},\ {F_i} \in {\mathbb{R}^{\delta_i \times {p_i}}},\ \text{and}\ G_i \in {\mathbb{R}^{\delta_i \times {p_i}}},
\end{equation*}
such that 
\begin{subequations}\label{BBTEF}
	\begin{align}
		T_{id}^{\top}{T_{id}} &= I_{\delta_i}\label{TTI}\\
        {G_i}{C_i}{B_{ - i}} &= T_{id}^{\top}{B_{ - i}} \label{GiCiB-i} \\
		{E_i}T_{id}^{\top} + ({F_i} - {E_i}{G_i}){C_i} &= (T_{id}^{\top} - {G_i}{C_i})A \\
		{\rm{Re}}\lambda ({E_i}) &< 0, \label{RelamE0}
	\end{align}
\end{subequations}
or such that \eqref{TTI} and
\begin{subequations}
	\begin{align}
		\delta_i &= p_i \tag{\ref{BBTEF}{e}}\label{deltaipi}\\
		E_i = F_i &= 0_{p_i \times p_i} \tag{\ref{BBTEF}{f}}\\
		G_i C_i &= T_{id}^\top. \tag{\ref{BBTEF}{g}}\label{GiCiTi}
	\end{align}
\end{subequations}
Based on \eqref{TTI}, there exists a matrix $T_{iu}$ such that 
\begin{equation}\label{Tiudef}
	T_{id}^\top T_{iu} = 0\ {\rm{and}}\ T_{iu}^\top T_{iu} = I_{n-\delta_i}.
\end{equation}
The above matrices\footnote{From the perspective of state-space decomposition, ${\rm{Im}} T_{id}$ and ${\rm{Im}} T_{iu}$ correspond to the locally detectable and undetectable subspaces, respectively. In other words, $T_{id}^\top x$ is reconstructible from $u_i$ and $y_i$, whereas $T_{iu}^\top x$ is not.} will be used in the observer design.

The following are dynamics of the $i$th observer node:
\begin{subequations}\label{obidynamic}
	\begin{align}
		{{\dot z}_i} &= {{\bar E}_i}{z_i} + {{\bar F}_i}{y_i} + {{\bar B}_i}{u_i} - {H_i}\left[ {\sum\nolimits_{j = 1}^N {{a_{ij}}({{\hat x}_i} - {{\hat x}_j})} } \right] \label{zidynamic}\\
		{{\hat x}_i} &= {z_i} + {{\bar G}_i}{y_i}, \label{hatxieq}
	\end{align}
\end{subequations}
where ${z}_i$, with initial value ${z}_i(0) = 0$, is an intermediate variable; ${\hat x}_i$ is the estimate of state $x$. The matrix gains in \eqref{obidynamic} are designed as follows:
\begin{subequations}\label{EFdesign}
	\begin{align}
		{{\bar E}_i} &= {T_{id}}{E_i}T_{id}^{\top} + {T_{iu}}T_{iu}^{\top}A \label{barE_i}\\
		{{\bar F}_i} &= {T_{id}}{F_i} + {T_{iu}}T_{iu}^{\top}A{{\bar G}_i} \label{barF_i}\\
		{{\bar G}_i} &= {T_{id}}{G_i} \label{barG_i}\\
		{{\bar B}_i} &= (I - {\bar G}_i{C_i}){B_i}. \label{barB_i}
	\end{align}
\end{subequations}
For notational simplicity, let $\varepsilon_{iu} =T_{iu}^\top \sum\nolimits_{j = 1}^N {a_{ij} (\hat x_i - \hat x_j)}$, and define a vector function ${h}(\cdot)$ as
\begin{equation*} \label{hfunc}
	{h}(\omega) = 
	\left\{
	\begin{aligned}
		\omega / \left\|\omega\right\|,\ \omega &\ne 0 \\
		0,\ \omega &= 0. 
	\end{aligned}
	\right.
\end{equation*}
The function $H_i(\cdot)$ in \eqref{obidynamic} is designed as follows:
\begin{equation} \label{Hdesign}
	{H_i}(\cdot) = {\gamma_i}{T_{iu}}\varepsilon_{iu} + {\gamma _{is}}{T_{iu}}{h}(\varepsilon_{iu}),
\end{equation}
where $\gamma_i$ and $\gamma_{is}$ are scalar gains evolving according to the following adaptive laws:
\begin{align*}
	{{\dot \gamma }_i} &= {\phi _i}{\left\| \varepsilon_{iu} \right\|^2} \\
	{{\dot \gamma }_{is}} &= {\phi _{is}}\left\| \varepsilon_{iu} \right\| 
\end{align*}
with step sizes ${\phi _{i}}$, ${\phi _{is}}$ and initial values $\gamma_i(0)$, $\gamma_{is}(0)$ chosen as positive constants. 
The convergence of the state estimation errors is guaranteed under the following assumptions:
\begin{assu}\label{spanwholespace}
Locally detectable subspaces span the whole space, i.e., $\sum\nolimits_{i = 1}^N {{\mathop{\rm Im}\nolimits} {T_{id}}} = \mathbb{R}^n$.
\end{assu}
\begin{assu}\label{ubound}
	There is a finite bound for $u_{-i}$ in \eqref{uiconstraint}, i.e., $\exists\bar u_{-}\in \mathbb{R},\ {\rm{s.t.}}\ \mathop {\max }\nolimits_{t\ge 0} \left\| u_{-i} (t) \right\| \le \bar u_{-},\ \forall i \in \mathcal{N}$.
\end{assu}
\begin{assu}\label{ooconnected}
	Communication graph $\mathcal{G}$ is connected.
\end{assu}
\noindent The following result comes from Theorem~1 in \cite{Ganghui2025arXiv}. 
\begin{lemma}\label{observerthm}\cite{Ganghui2025arXiv}
	Under Assumptions \ref{spanwholespace}-\ref{ooconnected}, the distributed observers with node dynamics \eqref{obidynamic} can produce accurate state estimates for system \eqref{LTIsys}, i.e.,
	\begin{equation*}
		\mathop {\lim }\nolimits_{t \to \infty } \left\| {{{\hat x}_i}(t) - x(t)} \right\| = 0,\ \forall i \in \mathcal{N}.
	\end{equation*}
	Moreover, adaptive gains $\gamma_i$ and $\gamma_{is}$ remain bounded, $\forall i \in \mathcal{N}$.
\end{lemma}
The remaining results of the present paper are developed based on this lemma and constitute new contributions.
\begin{rem}
In the presence of output measurement noise or other system uncertainties, the estimation error $\hat{x}_i - x$ cannot converge to zero exactly. As a result, the adaptive gains $\gamma_i$ and $\gamma_{is}$ will keep increasing. The following robust adaptive laws can address this problem:
\begin{align*}
    {{\dot \gamma }_i} &= -\sigma_i {\gamma }_i + {\phi_i}{\left\| \varepsilon_{iu} \right\|^2} \\
	{{\dot \gamma }_{is}} &= -\sigma_{is}{\gamma }_{i} + {\phi _{is}}\left\| \varepsilon_{iu} \right\|, 
\end{align*}
where $\sigma_{i}$ and $\sigma_{is}$ are positive scalars. The resulting estimation errors and adaptive gains will both remain bounded. Small $\sigma_{i}$ and $\sigma_{is}$, and large ${\phi_i}$ and ${\phi _{is}}$ help reduce the estimation errors. Such robust adaptive laws also apply to the distributed omniscient observers in Section~\ref{DOOD}.
\end{rem}

\begin{rem}
The methods of designing distributed observers without using global inputs were also investigated in \cite{G.Yang2022,G.Cao2023TAC,G.Cao2023Auto,G.Disaro2025}. However, the results therein rely on the following rank condition:
$$\operatorname{rank} (C_i B_{-i}) = \operatorname{rank} (B_{-i}),\ \forall i \in \mathcal{N},$$ 
which implicitly requires the local output dimension to be at least as large as the dimension of the locally unavailable input. Consequently, those results may not be suitable for solving the problems considered in Section~\ref{DOOD}, where 
$$\operatorname{rank} (C_i B_{-i}) < \operatorname{rank} (B_{-i}),\ \forall i \in \mathcal{N}.$$
\end{rem}

\section{Distributed Omniscient Observer Design}\label{DOOD}

In Section~\ref{RDOWUGI}, the revisited result only allows Assumption~\ref{spanwholespace} to be checked after each $T_{id}$ has been computed. Given that $T_{id}$ is coupled with other matrices in \eqref{BBTEF}, it is unclear for which class of systems Assumption~\ref{spanwholespace} can be satisfied. 
Moreover, the problem of simultaneously solving \eqref{BBTEF} and ensuring Assumption~\ref{spanwholespace} remains open.

In this Section, we show that linear MAS admit analytical solutions $\left\{T_{id},E_i,F_i,G_i\right\}_{i\in \mathcal{N}}$ that satisfy both \eqref{BBTEF} and Assumption~\ref{spanwholespace}. 
Then the distributed observers revisited in Section~\ref{RDOWUGI} can be directly deployed on each agent, enabling each agent to reconstruct the states of all agents. Such observers are referred to as the distributed omniscient observers in this paper, since they are specially designed for MAS. Precisely speaking, they are a special case of the revisited distributed observers.

\subsection{Design for Heterogeneous Multi-Agent Systems}\label{DHeMAS}

Consider a group of $N$ agents that have heterogeneous, general linear dynamics. The dynamics and output equation of the $i$th agent are described by
\begin{subequations}\label{HeMAS}
	\begin{align}
		\dot{\breve{x}}_i &= \breve{A}_i \breve{x}_i + \breve{B}_i u_i \\
		{y}_i &= \breve{C}_i \breve{x}_i,\ i \in \mathcal{N}, \label{heyidef} 
	\end{align}
\end{subequations}
where $\breve{x}_i \in \mathbb{R}^{n_i}$ is the state, $u_i \in \mathbb{R}^{m_i}$ is the control input, and $y_i \in \mathbb{R}^{p_i}$ is the measured output. The dynamics of the overall MAS take the same form as \eqref{LTIsys}, where $x = {\rm{col}}(\breve{x}_i)_{i = 1}^N$, $u = {\rm{col}}({u}_i)_{i = 1}^N$, $A = {\rm{diag}}(\breve{A}_i)_{i = 1}^N$, and $B = {\rm{diag}}(\breve{B}_i)_{i = 1}^N$. Moreover, \eqref{heyidef} aligns with \eqref{yidef}, where $C_i = \left[ {\begin{array}{*{20}{c}}
		{{0_{{p_i} \times \sum\nolimits_{q = 1}^{i - 1} {{n_q}} }}}&{\breve{C}_i}&{{0_{{p_i} \times \sum\nolimits_{q = i + 1}^N {{n_q}} }}}
\end{array}} \right]$.

Consider the case where each agent has access to the input and output of itself. Each agent serves as a node of the revisited distributed observers. For the $i$th agent, the matrices involved in \eqref{uiconstraint} are specified as follows:
	\begin{align}
		{B_i} &= {\left[ {\begin{array}{*{20}{c}}
					{{0_{{m_i} \times \sum\nolimits_{q = 1}^{i - 1} {{n_q}} }}}&{\breve{B}_i^{\top}}&{{0_{{m_i} \times \sum\nolimits_{q = i + 1}^N {{n_q}} }}}
			\end{array}} \right]^{\top}} \label{heteBidef}\\
		{B_{-i}} &= \left[\! {\begin{array}{*{20}{c}}
				\operatorname{diag}(\breve{B}_q)_{q = 1}^{i - 1}
				\!& \!0_{\sum_{q = 1}^{i - 1} \!\! n_q \times \sum_{q = i + 1}^{N} \!\! m_q} \\
				0_{n_i \times \sum_{q = 1}^{i - 1} \!\! m_q}
				\!& \!0_{n_i \times \sum_{q = i + 1}^{N} \!\! m_q} \\
				0_{\sum_{q = i + 1}^{N} \!\! n_q \times \sum_{q = 1}^{i - 1} \!\! m_q}
				\!& \!\operatorname{diag}(\breve{B}_q)_{q = i + 1}^{N}
		\end{array}} \!\right].\nonumber
	\end{align}
Accordingly, $u_{-i}$ in \eqref{uiconstraint} is formed by the inputs of all agents except agent $i$. If $(\breve{A}_i, \breve{C}_i)$ is detectable for all $i\in \mathcal{N}$, the analytical solutions that satisfy both \eqref{BBTEF} and Assumption~\ref{spanwholespace} are as follows:
\begin{subequations}\label{anasoluhe}
	\begin{align}
		{T_{id}} &= {\left[ {\begin{array}{*{20}{c}}
					{{0_{{n_i} \times \sum\nolimits_{q = 1}^{i - 1} {{n_q}} }}}&{{I_{{n_i}}}}&{{0_{{n_i} \times \sum\nolimits_{q = i + 1}^N {{n_q}} }}}
			\end{array}} \right]^{\top}} \\
		E_i &=  \breve{A}_i + \breve{L}_i \breve{C}_i,\ {F_i} = -{\breve{L}_i},\ G_i = 0_{n_i\times p_i},\ i \in \mathcal{N},
	\end{align}
\end{subequations}
where $\breve{L}_i$ is chosen such that $\breve{A}_i + \breve{L}_i \breve{C}_i$ has eigenvalues with negative real parts. 
Then the following result is straightforward from Lemma~\ref{observerthm}.

\begin{thm} \label{HeMASthm}
	Consider heterogeneous MAS \eqref{HeMAS}, where $(\breve{A}_i, \breve{C}_i)$ is detectable and $u_i$ is bounded, $\forall i \in \mathcal{N}$. 
	Suppose the agents can collect their own input $u_i$ and output $y_i$, and implement observer dynamics \eqref{obidynamic} over a connected communication graph $\mathcal{G}$. If the observers are designed according to \eqref{EFdesign}, \eqref{Hdesign}, \eqref{heteBidef}, and \eqref{anasoluhe}, then each agent can produce an accurate global state estimate, i.e.,
	\begin{equation*}
		\mathop {\lim }\nolimits_{t \to \infty } \left\| {{{\hat x}_i}(t) - x(t)} \right\| = 0,\ \forall i \in \mathcal{N}.
	\end{equation*}
	Moreover, adaptive gains $\gamma_i$ and $\gamma_{is}$ remain bounded, $\forall i \in \mathcal{N}$.	
\end{thm}
The distributed omniscient observers given in Theorem~\ref{HeMASthm} enable each agent to estimate the global state of MAS \eqref{HeMAS}, by only using its own input and output information while exchanging estimates with neighboring agents.

\subsection{Design for Homogeneous Multi-Agent Systems}\label{DHoMAS}
It follows directly that the distributed omniscient observers developed in Section~\ref{DHeMAS} can be applied to the following homogeneous MAS:
\begin{subequations}\label{HoMAS}
	\begin{align}
		\dot{\breve{x}}_i &= \breve{A} \breve{x}_i + \breve{B} \breve{u}_i \\
		\breve{y}_i &= \breve{C} \breve{x}_i,\ i \in \mathcal{N}, 
	\end{align}
\end{subequations}
where $\breve{x}_i \in \mathbb{R}^{\breve{n}}$ is the state, $\breve{u}_i \in \mathbb{R}^{\breve{m}}$ is the control input, and $\breve{y}_i \in \mathbb{R}^{\breve{p}}$ is the measured output. However, this requires each agent to measure its own output. For homogeneous MAS, the remainder of this section addresses the case where most agents can only measure the relative output between themselves and neighboring agents. 
Therefore, the results in Section~\ref{DHeMAS} cannot be considered as a more general case that encompasses the following results.

Let $\mathcal{R}$ denote a proper subset of $\mathcal{N}$, formed by the indices of those agents that have access to their own outputs. Accordingly, define the following two variables: 
\begin{subequations} \label{uycollect}
	\begin{align}
		{u}_i =&
		\begin{cases}
			\sum\nolimits_{j = 1}^N {{a_{ij}}(\breve{u}_i - \breve{u}_j)}, & i \in \mathcal{N} \setminus \mathcal{R} \\
			\ \ \ \ \ \ \ \ \ \breve{u}_i, & i \in \mathcal{R}
		\end{cases} \label{u_icollect}\\
		{y}_i =&
		\begin{cases}
			\sum\nolimits_{j = 1}^N {{a_{ij}}(\breve{y}_i - \breve{y}_j)}, & i \in \mathcal{N} \setminus \mathcal{R} \\
			\ \ \ \ \ \ \ \ \ \breve{y}_i, & i \in \mathcal{R},
		\end{cases}\label{yicaseho}
	\end{align}
\end{subequations}
which should be collected by the $i$th agent and be fed into observer dynamics \eqref{obidynamic}. 
Let ${\varpi_i} \in \mathbb{R}^N$ denote the $i$th standard basis vector, i.e., the $i$th column of $I_N$, and define the following row vector:
\begin{equation} \label{defbarL}
	\bar{\mathcal L}_i =
	\begin{cases}
		\varpi_i^{\top} \mathcal{L}, & i \in \mathcal{N} \setminus \mathcal{R} \\
		\ \varpi_i^{\top},   & i \in \mathcal{R}.
	\end{cases}
\end{equation}
The dynamics of the overall MAS take the same form as \eqref{LTIsys}, where $x = {\rm{col}}(\breve{x}_i)_{i = 1}^N$, $u = {\rm{col}}(\breve{u}_i)_{i = 1}^N$, $A = I_N \otimes \breve{A}$, and $B = I_N \otimes \breve{B}$. Moreover, \eqref{yicaseho} aligns with \eqref{yidef}, where $C_i = \breve{C} \left( {{\bar {\mathcal L}}_i} \otimes {I_{\breve{n}}} \right)$. 
Although the input available to the $i$th agent has been defined by \eqref{u_icollect}, the construction of $B_i$, $B_{-i}$, and $u_{-i}$ in \eqref{uiconstraint} is not as trivial as that in Section~\ref{DHeMAS}. Given that $B_{-i}$ should satisfy \eqref{GiCiB-i}, it has to be selected jointly with $T_{id}$. If $(\breve{A}, \breve{C})$ is detectable, we construct the following matrices:
\begin{subequations}\label{anasoluho}
	\begin{align}
		{B_i} &= \frac{{T_{id}}{\breve{B}}}{\left\| {{{\bar {\mathcal L}}_i}} \right\|},\ {B_{-i}} = T_{iu} \label{homoBiB-idef}\\
		T_{id}^\top &= \frac{{{\bar {\mathcal L}}_i} \otimes I_{\breve{n}}}{\left\| {{\bar {\mathcal L}}_i} \right\|}  \label{Tidho}\\
		E_i &= \breve{A} + \breve{L}_i \breve{C},\ G_i = 0_{\breve{n}\times \breve{p}}\\ 
		{F_i} &= -\frac{{\breve{L}_i}}{\left\| {{{\bar {\mathcal L}}_i}} \right\|},\ i \in \mathcal{N},
	\end{align}
\end{subequations}
where $T_{iu}$ is defined by \eqref{Tiudef}, ${\bar {\mathcal L}}_i$ is defined by \eqref{defbarL}, and $\breve{L}_i$ is chosen such that $\breve{A} + \breve{L}_i \breve{C}$ has eigenvalues
with negative real parts.
Before using Lemma~\ref{observerthm} to give the distributed omniscient observers, it suffices to examine if the construction in \eqref{anasoluho} can satisfy all conditions listed in Section~\ref{RDOWUGI}.
First, such construction is compatible with the relation \eqref{uiconstraint}, since there exists $u_{-i}$ satisfying \eqref{uiconstraint}, i.e.,
$${u_{-i}} = T_{iu}^\top \left( {I_N} \otimes \breve{B} \right)u.$$
Second, it can be verified that such construction satisfies \eqref{TTI}-\eqref{RelamE0}. 
Lastly, the following lemma ensures that Assumption~\ref{spanwholespace} always holds for such constructions.
\begin{lemma}\label{barLnonsin}
	If $\mathcal{L} \in \mathbb{R}^{N\times N}$ is a Laplacian matrix of a connected graph and $\mathcal{R} \neq \emptyset$, then $\bar{\mathcal{L}}_i$ defined in \eqref{defbarL} forms a nonsingular matrix ${\rm{col}}(\bar{\mathcal L}_i)_{i = 1}^N$.
\end{lemma}
The proof of Lemma~\ref{barLnonsin} is in Section~\ref{PLbarL}.
Based on Lemma~\ref{observerthm}, the following result is straightforward.

\begin{thm} \label{HoMASthm}
	Consider homogeneous MAS \eqref{HoMAS}, where $(\breve{A}, \breve{C})$ is detectable and $\breve{u}_i$ is bounded, $\forall i \in \mathcal{N}$. 
	Suppose the agents can collect input-output information as in \eqref{uycollect} and implement observer dynamics \eqref{obidynamic} over a connected communication graph $\mathcal{G}$. If $\mathcal{R} \neq \emptyset$ and the observers are designed according to \eqref{EFdesign}, \eqref{Hdesign}, and \eqref{anasoluho}, then each agent can produce an accurate global state estimate, i.e.,
	\begin{equation}\label{omniconverHo}
		\mathop {\lim }\nolimits_{t \to \infty } \left\| {{{\hat x}_i}(t) - x(t)} \right\| = 0,\ \forall i \in \mathcal{N}.
	\end{equation}
	Moreover, adaptive gains $\gamma_i$ and $\gamma_{is}$ remain bounded, $\forall i \in \mathcal{N}$.	
\end{thm}
Provided that there is at least one agent having access to its own output, the distributed omniscient observers given in Theorem~\ref{HoMASthm} enable each agent to estimate the global state of MAS \eqref{HoMAS} by using relative input-output information. 
Moreover, the observer gain design of the $i$th agent only relies on the dynamic model of the agents and the $i$th row vector of the Laplacian matrix; it does not require knowledge of the full Laplacian matrix.

\subsection{Further Extension}
The observers designed in Section~\ref{DHoMAS} require most agents to access the input information of their neighboring agents. This requirement may limit practical implementation when privacy, security, and communication constraints are considered. To remove this requirement, we develop a design method that does not require neighboring agents’ input information in this section. However, the design uses a scalar gain selected based on the global communication graph, and therefore no longer retains the fully distributed advantage of the results in Section~\ref{DHoMAS}.
Specifically, let us consider the distributed observers proposed in \cite{H.Zhang2011}:
\begin{equation}\label{HongweiObserver}
	\dot {\hat{\breve{x}}}_i = \breve{A} {\hat{\breve{x}}}_i + \breve{B} \breve{u}_i + cM\zeta_i,\ i \in \mathcal{N},
\end{equation}
where ${\hat{\breve{x}}}_i$ is the estimate of ${\breve{x}}_i$, and $\zeta_i$ is designed as
\begin{equation}\label{zetadef}
	\zeta_i = w_i \left({\breve{y}}_i - \breve{C}{\hat{\breve{x}}}_i \right) + {\sum\nolimits_{j = 1}^N {{a_{ij}}\left[{\breve{y}}_i - {\breve{y}}_j  - \breve{C}\left({\hat{\breve{x}}}_i - {\hat{\breve{x}}}_j \right) \right]} }
\end{equation}
with $w_{i}>0$ if the $i$th agent has access to its own output ${\breve{y}}_i$, and $w_{i}=0$ otherwise.
The scalar gain $c$ and matrix gain $M$ in \eqref{HongweiObserver} are designed as follows: 
\begin{equation}\label{cMdesign}
	c \ge \frac{1}{2 \lambda_{\min}\left(\mathcal{L}+W\right)},\ M = S\breve{C}^\top,
\end{equation} 
where $W = {\rm{diag}}(w_i)_{i=1}^N$, and $S$ is the unique positive definite solution of
\begin{equation*}
	\breve{A}S + S \breve{A}^\top - S \breve{C}^\top \breve{C}S + I = 0.
\end{equation*}
Based on \eqref{HongweiObserver}, the design of distributed omniscient observers \eqref{obidynamic} is given as follows:
\begin{subequations}\label{extengain}
	\begin{align}
		A &= I_N \otimes \breve{A},\ B_i = 0_{\breve{n}N \times 1},\ u_i = 0,\ y_i = {\hat{\breve{x}}}_i\\
		T_{id}^\top &= C_i = \begin{bmatrix}
			0_{\breve{n} \times (i-1)\breve{n}} & I_{\breve{n}} & 0_{\breve{n} \times (N-i)\breve{n}}
		\end{bmatrix}\\
		E_i &= F_i = 0_{\breve{n} \times \breve{n}},\ G_i = I_{\breve{n}},\ \forall i\in \mathcal{N}.
	\end{align}
\end{subequations}

\begin{thm} \label{extenHoMASthm}
	Consider homogeneous MAS \eqref{HoMAS}, where $(\breve{A}, \breve{C})$ is detectable and $\breve{u}_i$ is bounded, $\forall i \in \mathcal{N}$. 
	Suppose the agents implement observer dynamics \eqref{obidynamic} and \eqref{HongweiObserver} over a connected communication graph $\mathcal{G}$. If $W \neq 0$ and the observers are designed according to \eqref{EFdesign}, \eqref{Hdesign}, and \eqref{zetadef}-\eqref{extengain}, then each agent can produce an accurate global state estimate. Moreover, adaptive gains $\gamma_i$ and $\gamma_{is}$ remain bounded, $\forall i \in \mathcal{N}$.	
\end{thm}
The proof of Theorem~\ref{extenHoMASthm} is in Section~\ref{PoTexten}.

\section{Application A: Distributed Nash Equilibrium Seeking in Multi-Player Games} \label{AADNESMPG}

Consider a set of players indexed from $1$ to $N$. For each player $i \in \mathcal{N}$, let $\breve{x}_i \in \mathbb{R}^{\breve{n}}$ denote its action, and $J_i(x): \mathbb{R}^{\breve{n}N} \rightarrow \mathbb{R}$ denote its cost function, where $x = {\rm{col}}(\breve{x}_i)_{i=1}^N$. Define $\breve{x}_{-i} = \begin{bmatrix}
	\breve{x}_1^\top& \cdots & \breve{x}_{i-1}^\top & \breve{x}_{i+1}^\top &\cdots&\breve{x}_N^\top
\end{bmatrix}^\top$. The Nash equilibrium problem can be described as follows \cite{Guoqiang2022,Giuseppe2022}:
\begin{equation*}
	\min\nolimits_{\breve{x}_i \in \mathbb{R}^{\breve{n}}} J_i\left(\breve{x}_i,\breve{x}_{-i}\right),\ \forall i\in \mathcal{N}.
\end{equation*}
Accordingly, the Nash equilibrium refers to an action profile of all players $x^*={\rm{col}}\left(x_i^*\right)_{i=1}^N$ that satisfies 
\begin{equation*}
	J_i\left(x_i^*,x_{-i}^*\right) \le J_i\left(\breve{x}_i,x_{-i}^*\right),\ \forall \breve{x}_i \in \mathbb{R}^{\breve{n}},\ \forall i\in \mathcal{N}.
\end{equation*}
At the Nash equilibrium, no player can diminish its own cost by unilaterally changing its action. 

To solve the Nash equilibrium problem, define the game mapping as $\nabla J(x) = {\rm{col}}\left[\nabla_{\breve{x}_i}J_i(x)\right]_{i=1}^N$, provided that the cost function $J_i$ is continuously differentiable in $\breve{x}_i$.
The game mapping is said to be strongly monotone with constant $\mu >0$, if it holds that $\left(x_a - x_b\right)^\top\left[\nabla J(x_a) - \nabla J(x_b)\right] \ge \mu \left\|x_a - x_b\right\|^2$ for any $x_a,x_b \in \mathbb{R}^{\breve{n}N}$. 
The following is a basic centralized Nash equilibrium seeking algorithm:
\begin{equation} \label{bcnesl}
	\dot{\breve{x}}_i = - \nabla_{\breve{x}_i}J_i(x),\ \forall i \in \mathcal{N}.
\end{equation}
\begin{lemma}\label{lemmaneseek}\cite{Guoqiang2022}
	Suppose that each cost function $J_i\left(\breve{x}_i,\breve{x}_{-i}\right)$ is continuously differentiable and convex in $\breve{x}_i$ for every fixed $\breve{x}_{-i}$. If game mapping $\nabla J(x)$ is strongly monotone, then there exists a unique Nash equilibrium for the game and the trajectory of \eqref{bcnesl} converges to it, i.e., 
	\begin{equation*}
		\mathop {\lim }\nolimits_{t \to \infty } \left\| {{\breve{x}_i}(t) - x_i^*} \right\| = 0,\ \forall i \in \mathcal{N}.
	\end{equation*}
\end{lemma}

Algorithm \eqref{bcnesl} requires each player to have real-time access to global action profile $x$. 
For the sake of scalability, however, communications may only occur between neighboring agents in MAS.
In the case where only neighbors' actions are directly available, the distributed omniscient observers developed in Section~\ref{DOOD} can be used to implement algirithm \eqref{bcnesl}, by providing each player with an estimate of the global action profile. Specifically, the distributed Nash equilibrium seeking algorithm is designed as follows:
\begin{equation}\label{dnesa}
	\dot{\breve{x}}_i = {\breve{u}}_i,\ \forall i \in \mathcal{N},
\end{equation}
where ${\breve{u}}_i =  - \nabla_{\breve{x}_i}J_i(\hat{x}_i)$, and $\hat{x}_i$ comes from the distributed omniscient observers for MAS described by \eqref{dnesa} and $\breve{y}_i = \breve{x}_i$. See Theorem~\ref{HeMASthm}, Theorem~\ref{HoMASthm}, or Theorem~\ref{extenHoMASthm} for the observer design.
\begin{thm}\label{DNESAT}
	Suppose that each cost function $J_i\left(\breve{x}_i,\breve{x}_{-i}\right)$ is continuously differentiable and convex in $\breve{x}_i$ for every fixed $\breve{x}_{-i}$. Moreover, suppose that there exist two constants $\chi,\chi_s \ge 0$ such that
	\begin{equation}\label{relaxassum}
		\left\|\nabla_{\breve{x}_i} J_i(x_a) - \nabla_{\breve{x}_i} J_i(x_b)\right\|^2 \le \chi\left\| x_a - x_b \right\|^2 + \chi_s \left\| x_a - x_b \right\|,
	\end{equation}
	$\forall x_a,x_b \in \mathbb{R}^{\breve{n}N},\ \forall i \in \mathcal{N}$. If the game mapping $\nabla J(x)$ is strongly monotone, then implementing algorithm \eqref{dnesa} based on the observers presented in Theorem~\ref{HeMASthm}, Theorem~\ref{HoMASthm}, or Theorem~\ref{extenHoMASthm} gives the unique Nash equilibrium of the game, i.e., 
	\begin{equation*}
		\mathop {\lim }\nolimits_{t \to \infty } \left\| {{\breve{x}_i}(t) - x_i^*} \right\| = 0,\ \forall i \in \mathcal{N}.
	\end{equation*}
\end{thm}
See Section~\ref{PTDNESAT} for the proof of Theorem~\ref{DNESAT}. Further discussions on Theorem~\ref{DNESAT} are as follows:
\begin{itemize}
	\item Condition \eqref{relaxassum} is more general than the following Lipschitz condition: There exists a constant $\bar{\chi} \ge 0$ such that
	\begin{equation}\label{origassum}
		\left\|\nabla_{\breve{x}_i} J_i(x_a) - \nabla_{\breve{x}_i} J_i(x_b)\right\| \le \bar{\chi}\left\| x_a - x_b \right\|,
	\end{equation}
	$\forall x_a,x_b \in \mathbb{R}^{\breve{n}N},\ \forall i \in \mathcal{N}$. For example, for scalar $\breve{x}_i$, 
	\begin{equation*}
		\nabla_{\breve{x}_i} J_i(x)=
		\begin{cases}
			\breve{x}_i+\sqrt{\breve{x}_i}, & \breve{x}_i\ge0\\
			\breve{x}_i-\sqrt{-\breve{x}_i}, & \breve{x}_i<0
		\end{cases}
	\end{equation*}
	satisfies \eqref{relaxassum}, while it does not satisfy \eqref{origassum}.
	\item According to Theorem~\ref{HeMASthm} and Theorem~\ref{HoMASthm}, if ${\breve{u}}_i$ in \eqref{dnesa} is bounded\footnote{Taking ${\breve{u}}_i =  - \nabla_{\breve{x}_i}J_i(\hat{x}_i)$ as an example, the boundedness can be guaranteed by assuming that each player's action belongs to a bounded closed subset of $\mathbb{R}^{\breve{n}}$, and $\nabla_{\breve{x}_i}J_i$ satisfies Lipschitz condition on this subset with a Lipschitz extension \cite{J.Heinonen2005} outside this subset.}, the distributed omniscient observers can fulfill \eqref{omniconverHo}, which does not rely on the specific value of ${\breve{u}}_i$. This implies that the seeking algorithm and the distributed omniscient observers can be designed separately. The separability may help accommodate a variety of seeking algorithms for the solution of more complex Nash equilibrium problems in future research.
\end{itemize}

\section{Application B: Self-Organized Social Behavior Emergence in Artificial Swarms} \label{ABSSBEAS}

Two bio-inspired simulation examples in this section demonstrate possible use cases of the proposed distributed omniscient observers.
Since there is no command center coordinating the agents, the following decision and action mechanism is referred to as a self-organized way to bring out collective intelligent behaviors of them.

\subsection{Confine Companions to a Convex Hull} \label{CCTACH}

The first example is inspired by the herding behaviors of sheepdogs\,---\,they collaborate with each other to gather and move livestock from one place to another. 
In this example, there are leader agents and follower agents.
The leaders can move freely, which represents the behavior of herding sheepdogs.
The followers will identify which agents are leaders and assemble into the convex hull formed by the leaders. 

\textbf{Basic Setup:} Within an x-y plane, dynamics of the agents indexed from $1$ to $N$ are of the form \eqref{HoMAS}, where
\begin{align*}
	{\breve{x}}_i =& \begin{bmatrix}
		{\breve{p}}_{i}^x \\
		{\breve{p}}_{i}^y \\
		{\breve{z}}_{i}
	\end{bmatrix},\ {\breve{A}}	= \begin{bmatrix}
		0&0&0\\
		0&0&0\\
		0&0&0
	\end{bmatrix},\ 
	{\breve{B}}	= \begin{bmatrix}
		1&0&0\\
		0&1&0\\
		0&0&1
	\end{bmatrix},\\ {\breve{u}}_i =& \begin{bmatrix}
		{\breve{v}}_{i}^x \\
		{\breve{v}}_{i}^y \\
		{\breve{u}}_{i}^z
	\end{bmatrix},\ 
	{\breve{C}} = \begin{bmatrix}
		1&0&0\\
		0&1&0\\
		0&0&1
	\end{bmatrix},
\end{align*}
with ${\breve{p}}_{i}^x$ and ${\breve{p}}_{i}^y$ denoting the positions, ${\breve{v}}_{i}^x$ and ${\breve{v}}_{i}^y$ the velocities, ${\breve{z}}_{i}$ the identity state, and ${\breve{u}}_{i}^z$ the identity input of the $i$th agent.
Based on a connected graph $\mathcal{G}$ and a nonempty set $\mathcal{R}$, the agents collect relative/absolute input-output information, i.e., $u_i$ and $y_i$ defined in \eqref{uycollect}, and carry out the distributed omniscient observers presented in Theorem~\ref{HoMASthm}.

\textbf{Leaders' Actions:} The velocities of leaders are freely chosen\footnote{The trajectories of leaders can be designed by choosing their velocities, which can be used to guide followers through obstacles, or to serve other practical purposes.}. The identity input of a leader is chosen as ${\breve{u}}_{j}^z = - {\breve{z}}_{j} + z^*$, where $j$ is the leader agent's index and $z^*$ is a positive constant. The identity input is used to increase the identity state, so that followers can tell which agents are leaders according to the estimated identity states produced by the distributed omniscient observers.

\textbf{Followers' Decisions:}
Each follower determines in real time a set of candidate leaders for itself, based on the estimates of the identity states of all the other agents.
From the perspective of a follower agent, anyone of the other agents will be labeled as a candidate leader, if the estimated identity state of the agent is greater than $z^*_t$, a positive threshold chosen to be lower than $z^*$.

\textbf{Followers' Actions:} 
Each follower heads toward a candidate target point, that is a convex combination of the estimated positions of the candidate leaders.
A follower with index $k$ will use the estimate for ${\breve{x}}_k$ provided in $\hat{x}_k$ to design control inputs ${\breve{v}}_{i}^x$ and ${\breve{v}}_{i}^y$ for itself.
The identity input of the follower is chosen as ${\breve{u}}_{k}^z = - {\breve{z}}_{k}$.

\begin{figure*}[htpb]\centering
	\centerline{\includegraphics[width=0.94\textwidth]{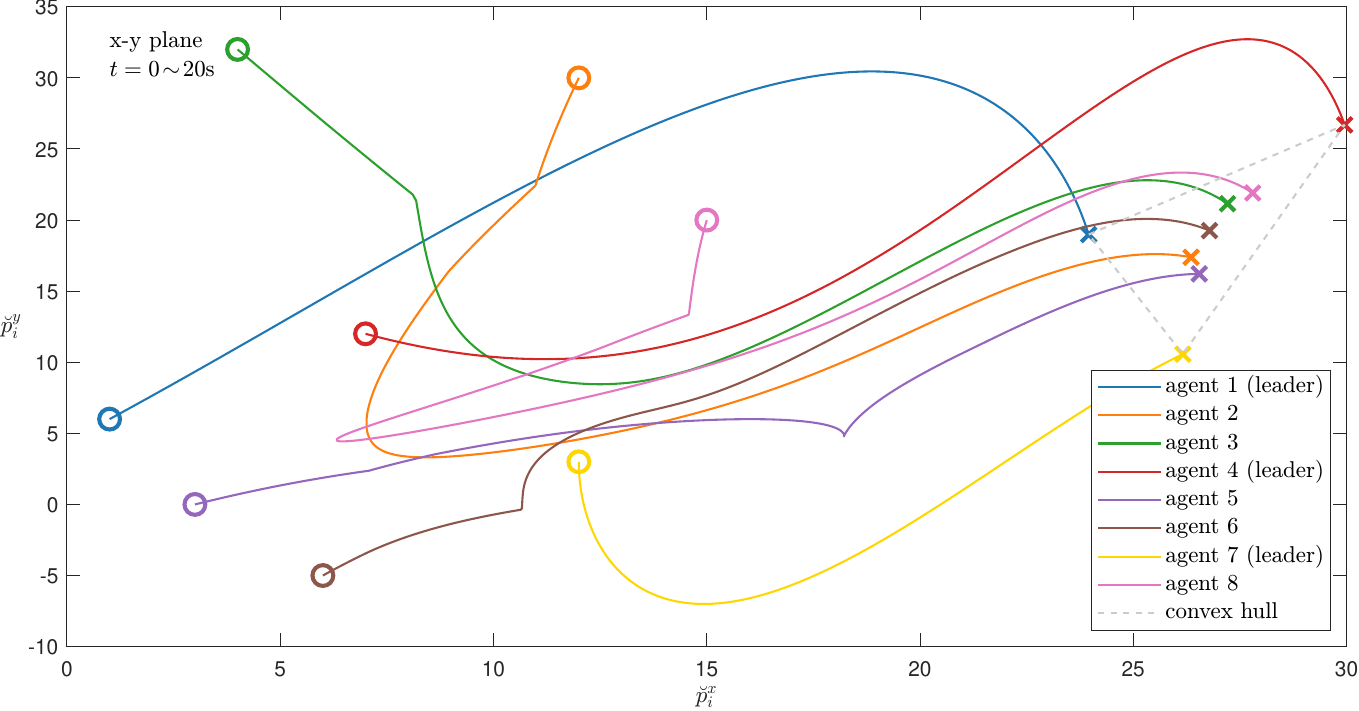}}
	\caption{Agent trajectories (start/end positions denoted by circles/crosses) in Section~\ref{CCTACH}.}
	\label{trajectories}
\end{figure*}
\begin{figure*}[htpb]\centering
	\centerline{\includegraphics[width=1\textwidth]{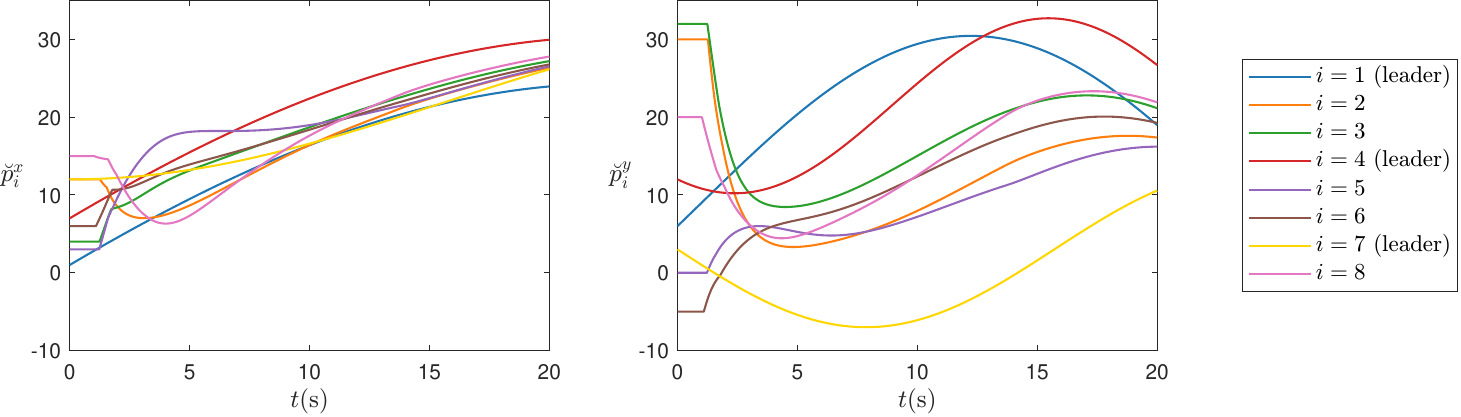}}
	\caption{Agent positions vs. time in Section~\ref{CCTACH}.}
	\label{pvst_sheepdog}
\end{figure*}
\begin{figure*}[htpb]\centering
	\centerline{\includegraphics[width=1\textwidth]{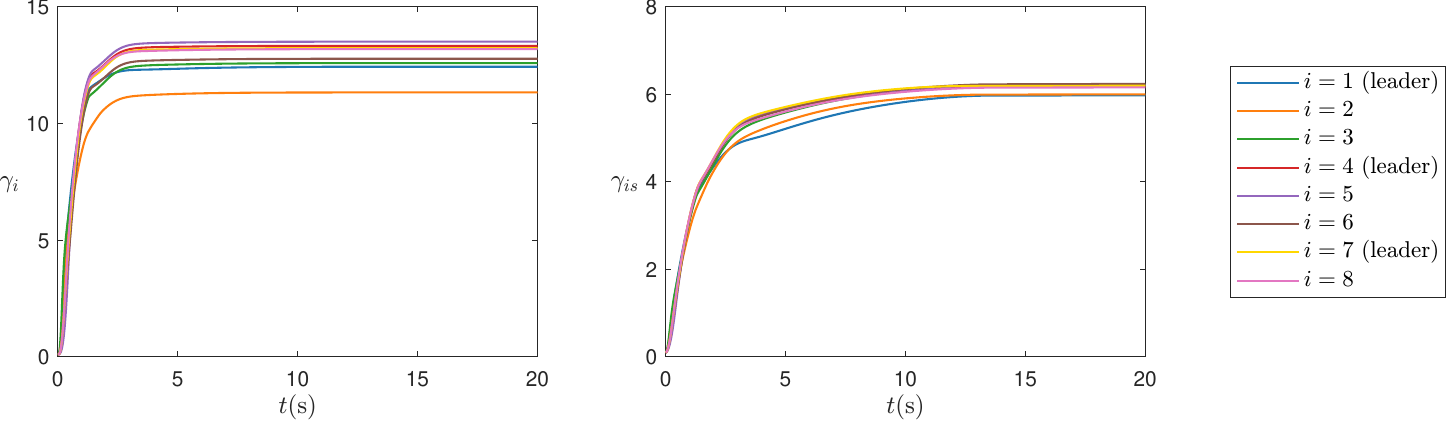}}
	\caption{Adaptive gains vs. time in Section~\ref{CCTACH}.}
	\label{gammavst_sheepdog}
\end{figure*}
\begin{figure*}[htpb]\centering
	\centerline{\includegraphics[width=0.88\textwidth]{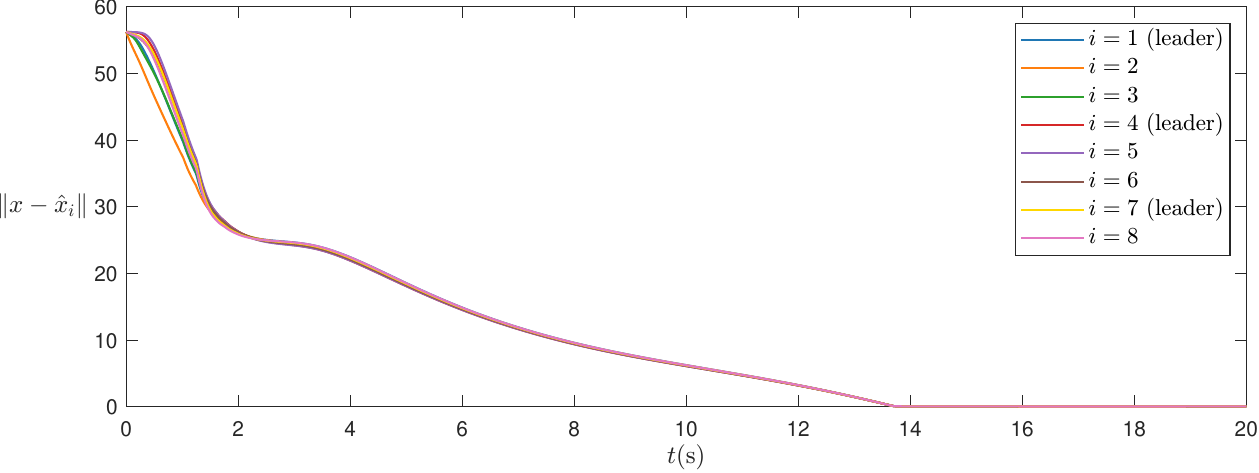}}
	\caption{Estimation error norms vs. time in Section~\ref{CCTACH}, where $x = {\rm{col}}\left(\breve{x}_i\right)_{i=1}^{8}$.}
	\label{evst_sheepdog}
\end{figure*}

\subsection{Summon Companions by Circling}\label{SCBC}

The second example is inspired by the dancing behaviors of honeybees\,---\,they use dance language to communicate the location and the abundance of nectar sources to other members of the hive. In this example, the leader agents will circle at different places with different speeds, which emulates the behavior of the dancing honeybees.
The follower agents will assemble at the leaders' places respectively, according to the ratio of their circling speeds. Followers tend to be attracted to leaders with higher speeds.

\textbf{Basic Setup:} Within an x-y plane, dynamics of the agents indexed from $1$ to $N$ are of the form \eqref{HoMAS}, where
\begin{align*}
	{\breve{x}}_i =& \begin{bmatrix}
		{\breve{p}}_{i}^x \\
		{\breve{p}}_{i}^y \\
		{\breve{v}}_{i}^x \\
		{\breve{v}}_{i}^y \\
	\end{bmatrix},\ {\breve{A}}	= \begin{bmatrix}
		0&0&1&0\\
		0&0&0&1\\
		0&0&0&0\\
		0&0&0&0
	\end{bmatrix},\ 
	{\breve{B}}	= \begin{bmatrix}
		0&0\\
		0&0\\
		1&0\\
		0&1
	\end{bmatrix},\\ {\breve{u}}_i =& \begin{bmatrix}
		\breve{a}_i^x\\
		\breve{a}_i^y
	\end{bmatrix},\ 
	{\breve{C}} = \begin{bmatrix}
		1&0&0&0\\
		0&1&0&0
	\end{bmatrix},
\end{align*}
with ${\breve{p}}_{i}^x$ and ${\breve{p}}_{i}^y$ denoting the positions, ${\breve{v}}_{i}^x$ and ${\breve{v}}_{i}^y$ the velocities, and $\breve{a}_i^x$ and $\breve{a}_i^y$ the accelerations of the $i$th agent.
Based on a connected graph $\mathcal{G}$ and a nonempty set $\mathcal{R}$, the agents collect relative/absolute input-output information, i.e., $u_i$ and $y_i$ defined in \eqref{uycollect}, and carry out the distributed omniscient observers presented in Theorem~\ref{HoMASthm}.

\textbf{Leaders' Actions:} 
Leaders control their movements according to the estimated positions and velocities of themselves. In other words, a leader with index $j$ will use the estimate for ${\breve{x}}_j$ provided in $\hat{x}_j$ to design control input ${\breve{u}}_j$ for itself.
If the leader intends to attract followers to position $(\bar{p}^x_j,\bar{p}^y_j)$, it will control itself to asymptotically move at a speed $v^*_j$ anticlockwise along the circumference of a circle centered at $(\bar{p}^x_j,\bar{p}^y_j)$ with a unit radius.

\textbf{Followers' Decisions:} 
Each follower determines in real time a candidate leader for itself, based on the estimates of the positions and velocities of all agents produced by the distributed omniscient observers.
From the perspective of a follower agent, each of other agents will be labeled as a candidate leader/follower, if the estimated speed (the Euclidean norm of the estimated velocity vector) of the agent is greater/less than ${v}^*_t$, a positive threshold chosen to be lower than the minimum circling speed of the leaders.
Then, the follower will assign each candidate follower (including itself) a candidate leader that is as near as possible, such that the number of each candidate leader's candidate followers is in proportion to the estimates of the candidate leaders' speeds.

\textbf{Followers' Actions:} Each follower heads toward a candidate target point indicated by its candidate leader. The point is one unit away from the estimated position of the candidate leader, along the direction indicated by a $\pi/2$ anticlockwise rotation of the estimated velocity vector of the candidate leader. A follower with index $k$ will use the estimate for ${\breve{x}}_k$ provided in $\hat{x}_k$ to design control input ${\breve{u}}_k$ for itself. 

\begin{figure*}[htpb]\centering	
	\centerline{\includegraphics[width=1\textwidth]{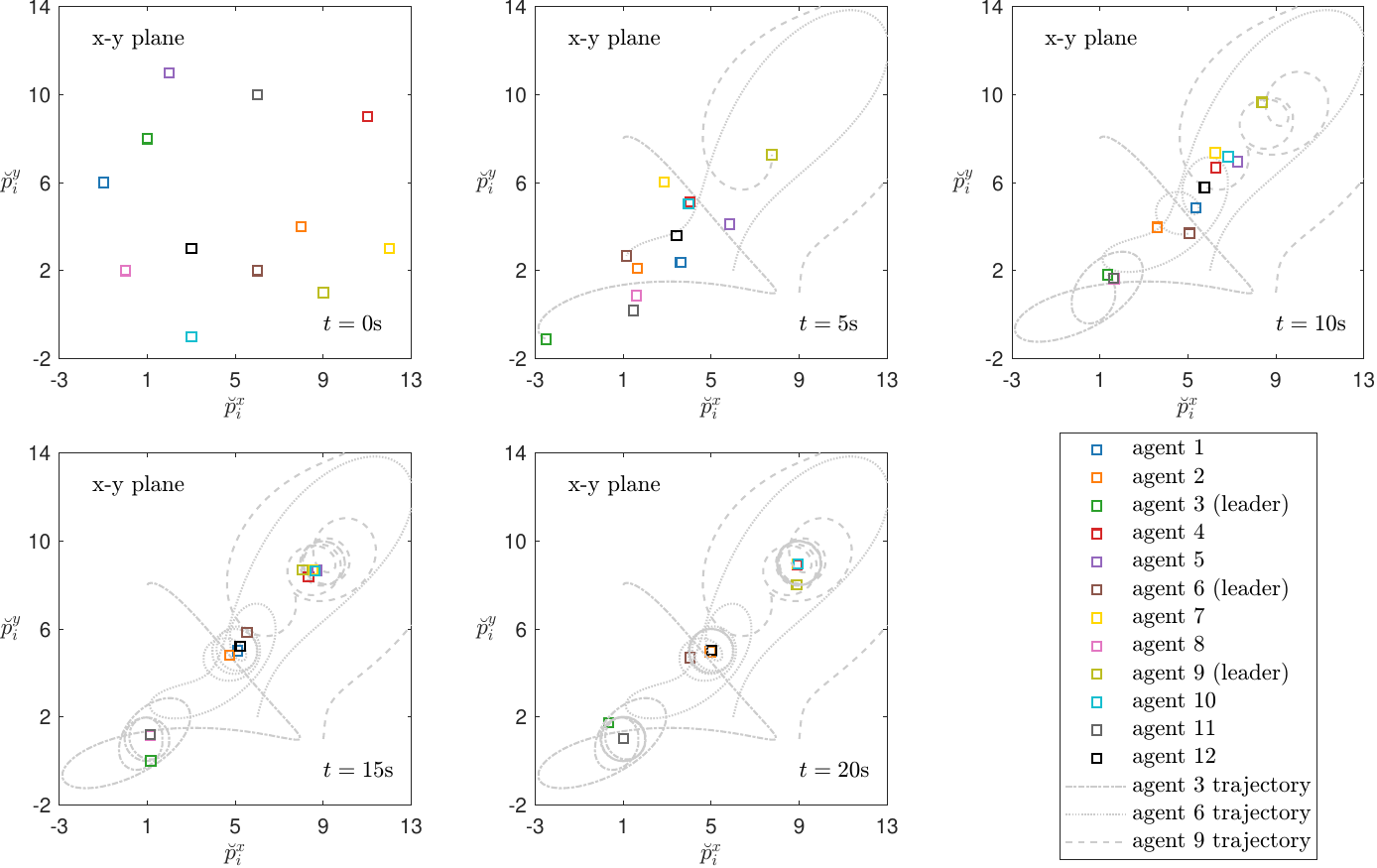}}
	\caption{Snapshots of agent positions in Section~\ref{SCBC}.}
	\label{snapshots}
\end{figure*}
\begin{figure*}[htpb]\centering
	\centerline{\includegraphics[width=1\textwidth]{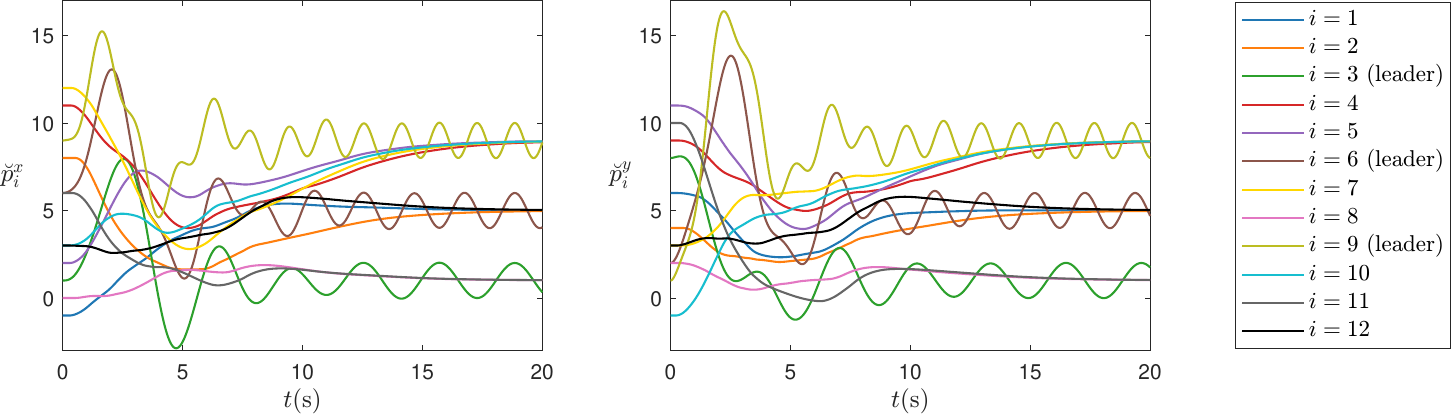}}
	\caption{Agent positions vs. time in Section~\ref{SCBC}.}
	\label{pvst_bee}
\end{figure*}
\begin{figure*}[htpb]\centering
	\centerline{\includegraphics[width=1\textwidth]{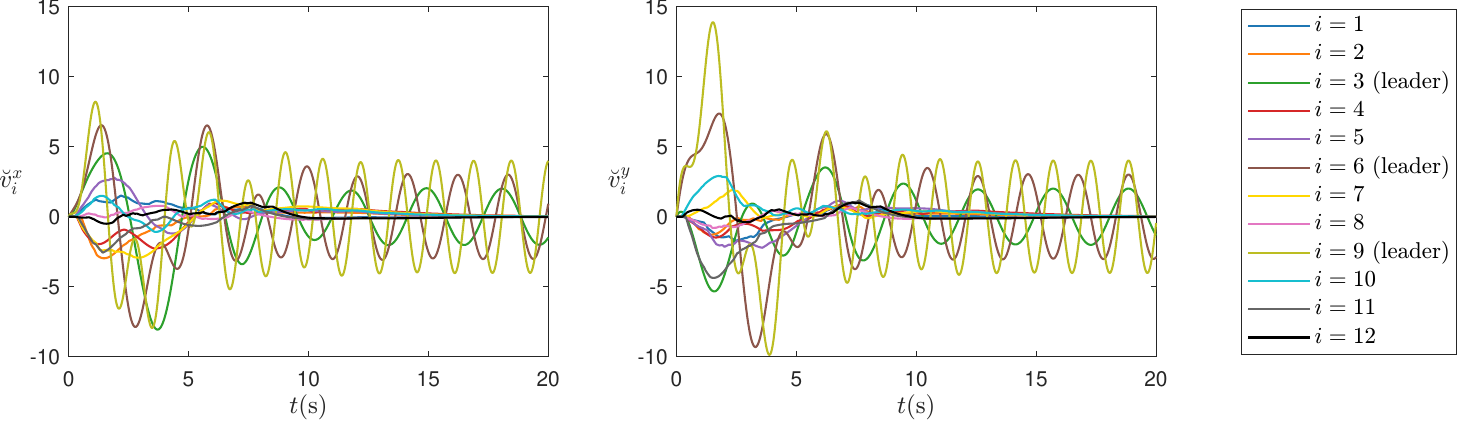}}
	\caption{Agent velocities vs. time in Section~\ref{SCBC}.}
	\label{vvst_bee}
\end{figure*}
\begin{figure*}[htpb]\centering
	\centerline{\includegraphics[width=1\textwidth]{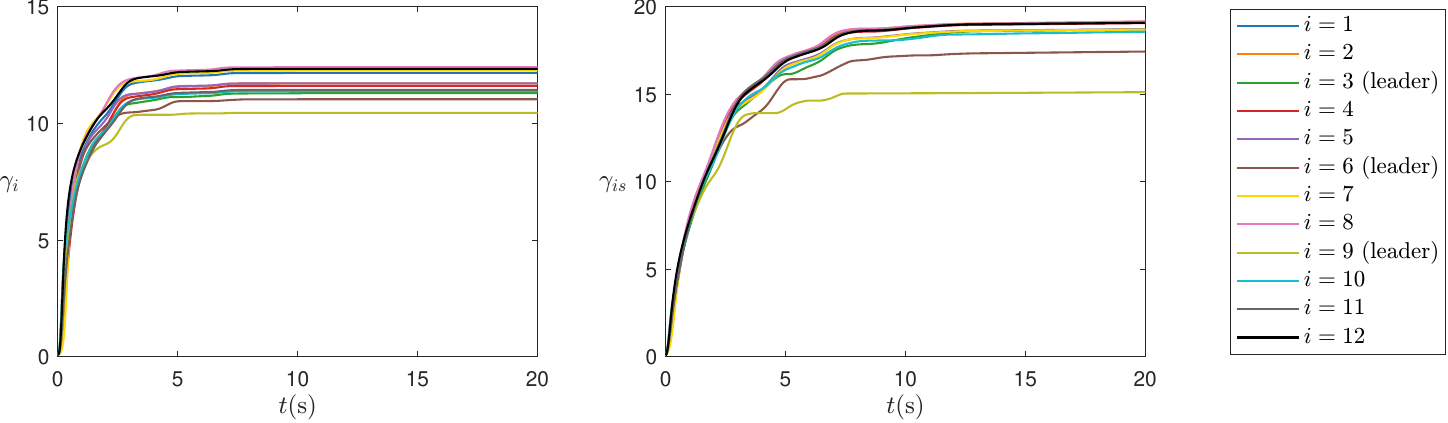}}
	\caption{Adaptive gains vs. time in Section~\ref{SCBC}.}
	\label{gammavst_bee}
\end{figure*}
\begin{figure*}[htpb]\centering
	\centerline{\includegraphics[width=0.88\textwidth]{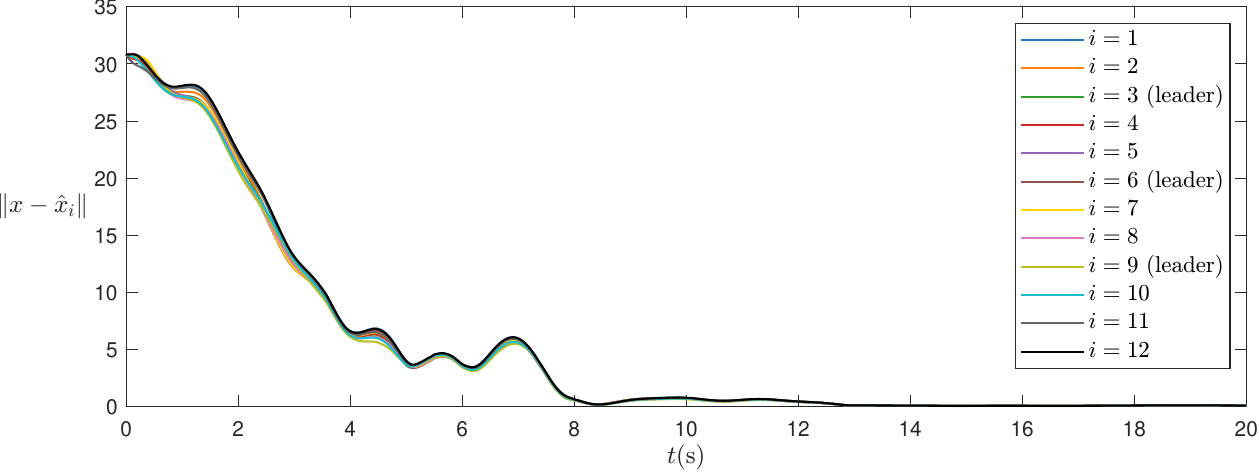}}
	\caption{Estimation error norms vs. time in Section~\ref{SCBC}, where $x = {\rm{col}}\left(\breve{x}_i\right)_{i=1}^{12}$.}
	\label{evst_bee}
\end{figure*}

\subsection{Simulation Results and Discussions}
Figs.~\ref{trajectories}-\ref{evst_bee} demonstrate the simulation results of the two examples formulated in Section~\ref{CCTACH} and \ref{SCBC}, respectively. 
In the first example, agents 2 and 7 have access to their own absolute positions. In the second example, agents 2, 8, and 11 have access to their own absolute positions; the three leaders' circling speeds are designed to be $v^*_3 = 2$, $v^*_6 = 3$, and $v^*_9 = 4$. 
Other parameter settings are specified as follows. For the first example, 
\begin{align*}
\mathcal{L} &=
\begin{bmatrix}
2 & -1 & 0 & 0 & 0 & 0 & 0 & -1 \\
-1 & 2 & -1 & 0 & 0 & 0 & 0 & 0 \\
0 & -1 & 3 & -1 & 0 & -1 & 0 & 0 \\
0 & 0 & -1 & 2 & -1 & 0 & 0 & 0 \\
0 & 0 & 0 & -1 & 2 & -1 & 0 & 0 \\
0 & 0 & -1 & 0 & -1 & 3 & -1 & 0 \\
0 & 0 & 0 & 0 & 0 & -1 & 2 & -1 \\
-1 & 0 & 0 & 0 & 0 & 0 & -1 & 2
\end{bmatrix}\\
{\phi _{i}} &= {\phi _{is}} = 1,\ \gamma_i(0) = \gamma_{is}(0) = 0.1\\
{\breve{L}_i} &= \begin{bmatrix}
-1.2198 & 0 & 0 \\
0 & -1.2198 & 0 \\
0 & 0 & -1.2198
\end{bmatrix},\ i= 1,2,\ \cdots,8.
\end{align*}
For the second example, 
\begin{align*}
\mathcal{L} &=
\begin{bmatrix}
2&-1&0&0&0&0&0&0&0&0&0&-1\\
-1&3&-1&0&0&-1&0&0&0&0&0&0\\
0&-1&3&-1&0&0&0&0&0&-1&0&0\\
0&0&-1&3&-1&0&0&-1&0&0&0&0\\
0&0&0&-1&2&-1&0&0&0&0&0&0\\
0&-1&0&0&-1&3&-1&0&0&0&0&0\\
0&0&0&0&0&-1&2&-1&0&0&0&0\\
0&0&0&-1&0&0&-1&3&-1&0&0&0\\
0&0&0&0&0&0&0&-1&2&-1&0&0\\
0&0&-1&0&0&0&0&0&-1&3&-1&0\\
0&0&0&0&0&0&0&0&0&-1&2&-1\\
-1&0&0&0&0&0&0&0&0&0&-1&2
\end{bmatrix}\\
{\phi _{i}} &= 2,\ {\phi _{is}} = 4,\ \gamma_i(0) = \gamma_{is}(0) = 0.1\\
{\breve{L}_i} &=
\begin{bmatrix}
-2.1667 & 0 \\
0 & -2.1667 \\
-1.4139 & 0 \\
0 & -1.4139
\end{bmatrix},\ i=1,2,\ \cdots,12.
\end{align*}

The following are several interpretations and discussions of the simulation results.
\begin{itemize}
	\item In both examples, just a few agents have access to their own positions. Other agents only access relative position information. Moreover, the agents measure neither the absolute nor the relative velocity information.
	\item Figs.~\ref{trajectories} and \ref{pvst_sheepdog} show that all followers can move into the convex hull formed by the leaders. Since the leaders are allowed to move freely, the shape and position of the convex hull can be time-varying.
	\item Fig.~\ref{snapshots} shows that the leaders wiggle before circling. This is because they use the state estimate, not the true state, to control themselves.
	\item Figs.~\ref{pvst_bee} and \ref{vvst_bee} show that two, three, and four followers are attracted to the three leaders respectively, in proportion to the circling speeds of the leaders. Different initial positions of the agents may lead to a different final result, but will preserve the ratio of the number of followers that each leader attracts.
	\item In both examples, each follower can identify which agents are the leaders autonomously. This is achieved by estimating the augmented identity state (in the first example) and the velocity (in the second example) of agents.
    \item Fig.~\ref{pvst_sheepdog} shows that the followers keep still at the beginning. This is because the identity states of the leaders have not reached threshold $z^*_t$, and the state estimates have not reached a desired level of accuracy. Therefore, the followers cannot identify the leaders correctly in the first few seconds.
	\item The leader-identification mechanisms can accommodate the change of leaders. Suppose in the second example that agents 3 and 6 keep circling at their speeds, while agent 9 stops circling after $t=20$s. Then, for agent 9 and its four followers, two of them will head toward position $(1,1)$, and the other three will go to $(5,5)$ autonomously.
    \item Figs. \ref{evst_sheepdog} and \ref{evst_bee} show that the state estimation errors of the distributed omniscient observers converge to zero, even though the inputs of agents are persistently nonzero. Figs. \ref{gammavst_sheepdog} and \ref{gammavst_bee} show that the adaptive gains in the distributed omniscient observers remain bounded.
    \item Estimating the global states of MAS incurs computational and communication loads that scale with both the number of agents $N$ and the state dimension of each agent $\breve{n}$. Specifically, each agent exchanges an $\breve{n}N$-dimensional vector with neighboring agents and runs $\breve{n}N$-dimensional observer dynamics.
\end{itemize}

\section{Conclusion} \label{Conclusion}
In this paper, distributed omniscient observers are proposed for heterogeneous and homogeneous linear MAS, respectively. 
The observer design for the latter is based on mostly relative, as well as a small amount of absolute, local input-output information of the agents. 
Knowledge of the global communication graph can obviate the relative input information required for the design, and vice versa.
As a result, each agent can estimate the states of itself and other agents. The state estimation errors can converge to zero even though the inputs of other agents are persistently nonzero. 
An application in distributed Nash equilibrium seeking, and two bio-inspired simulation examples show that the proposed distributed omniscient observers can contribute to the emergence of collective intelligence in MAS.

Since the computational and communication resources required for global state estimation in MAS grow with the number of agents, partitioning agents into smaller groups provides a scalable solution for large-scale MAS. In this way, distributed omniscient observers can be designed within each group to support intra-group cooperation while reducing the dimension of the exchanged information and observer dynamics. Effective inter-group coordination strategies remain to be investigated in future work. In addition, how to improve the state estimation accuracy in the presence of output measurement noise is another important direction for future research.

\section{Appendix} \label{append}
\subsection{Proof of Lemma~\ref{barLnonsin}}\label{PLbarL}
First, consider the case where set $\mathcal{R}$ contains exactly one element $r$. There exists a permutation operation that rearranges the $r$th row and column in matrix ${\rm{col}}(\bar{\mathcal L}_i)_{i = 1}^N$ to the bottom and the rightmost respectively, i.e.,
\begin{equation*}
	W^\top {\rm{col}}(\bar{\mathcal L}_i)_{i = 1}^N W = \begin{bmatrix}
		{\mathcal L}^W_{1,1} & *\\
		0 & 1
	\end{bmatrix},
\end{equation*}
where $W$ is a permutation matrix, and ${\mathcal L}^W_{1,1}$ is a nonsingular matrix according to Lemma 5 in \cite{Ganghui25TAC}. It follows that ${\rm{col}}(\bar{\mathcal L}_i)_{i = 1}^N$ is also nonsingular. For the case where set $\mathcal{R}$ contains more than one element, similar analysis can be done.

\subsection{Proof of Theorem~\ref{extenHoMASthm}} \label{PoTexten}
It follows from \cite{H.Zhang2011} that $\left\| {{{\hat{\breve{x}}}_i}(t) -{\breve{x}}_i(t)} \right\|$, i.e., $$\left\|T_{id}^\top \left( \hat{x}_i(t)-x(t) \right)\right\|,$$ exponentially converges to zero, $\forall i \in \mathcal{N}$. Then it can be proved as in the proof of Theorem~1 in \cite{Ganghui2025arXiv} that adaptive gains remain bounded and $\mathop {\lim }\nolimits_{t \to \infty } \left\|T_{iu}^\top \left( \hat{x}_i(t)-x(t) \right) \right\| = 0,\ \forall i \in \mathcal{N}$. 

\subsection{Proof of Theorem~\ref{DNESAT}} \label{PTDNESAT}
Consider the following Lyapunov candidate function
\begin{align}
	{V} =\frac{1}{2}\left(x - x^*\right)^\top \left(x - x^*\right) + V_o,  \label{Vlabel}
\end{align}
where $V_o$ is a Lyapunov function constructed in \cite{Ganghui2025arXiv}, that is
\begin{align}
	V_o &= \frac{1}{2}\varepsilon _u^{\top}{\left[ {T_u^{\top}({\mathcal L} \otimes I){T_u}} \right]^{ - 1}}{\varepsilon _u} + \sum\nolimits_{i = 1}^N {\frac{1}{{2{\phi _i}}}{{({\gamma _i} - {\gamma ^*})}^2}} \nonumber \\
	&\ \ \ \ + \sum\nolimits_{i = 1}^N {\frac{1}{{2{\phi _{is}}}}{{({\gamma _{is}} - \gamma_s^*)}^2}}. \label{Volabel}
\end{align}
In \eqref{Volabel}, $\varepsilon_u = {\rm col}\left(\varepsilon_{iu}\right)_{i=1}^N$, $\varepsilon_{iu} =T_{iu}^\top \sum\nolimits_{j = 1}^N {a_{ij} (\hat x_i - \hat x_j)}$, $T_u = {\rm col}\left(T_{iu}\right)_{i=1}^N$, and $\gamma^*$ and $\gamma_s^*$ are two positive constants to be determined. Differentiating \eqref{Vlabel} along the trajectory of \eqref{dnesa} yields that   
\begin{align*}
	\dot V =& -\left(x - x^*\right)^\top {\rm{col}}\left[\nabla_{\breve{x}_i}J_i(x) + \nabla_{\breve{x}_i}J_i(\hat{x}_i) - \nabla_{\breve{x}_i}J_i(x) \right]_{i=1}^N \\
	&+ \dot{V}_o \\
	\le& -\mu \left\|x - x^*\right\|^2 \!+ \left(x - x^*\right)^\top\!\! {\rm{col}}\!\left[\nabla_{\breve{x}_i}J_i(x) - \nabla_{\breve{x}_i}J_i(\hat{x}_i) \right]_{i=1}^N \\
	&+ \dot{V}_o\\
	\le& -\frac{\mu}{2} \left\|x - x^*\right\|^2 + 
	\frac{1}{2\mu} \left\|{\rm{col}}\left[\nabla_{\breve{x}_i}J_i(x) - \nabla_{\breve{x}_i}J_i(\hat{x}_i) \right]_{i=1}^N\right\|^2 \\
	&+ \dot{V}_o\\
	\le& -\frac{\mu}{2} \left\|x - x^*\right\|^2 + V_o^\prime,
\end{align*}
where $V_o^\prime = \frac{\chi}{2\mu}\sum\nolimits_{i = 1}^N \left\| x - \hat{x}_i \right\|^2 + \frac{\chi_s}{2\mu} \sum\nolimits_{i = 1}^N \left\| x - \hat{x}_i \right\|+ \dot{V}_o$. It can be verified, by using the same analysis method as in \cite{Ganghui2025arXiv}, that there exist four positive constants $\gamma^*$, $\gamma_s^*$, $\lambda^*$, and $\lambda^*_s$, such that $V_o^\prime \le \lambda^* \left\| {{\varepsilon _d}} \right\|^2 + \lambda^*_s \left\| {{\varepsilon _d}} \right\|$, where $\varepsilon_d = {\rm col}\left(\varepsilon_{id}\right)_{i=1}^N$ and ${\varepsilon _{id}} = T_{id}^{\top}\left({\hat x}_i - x\right)$. Therefore, 
\begin{equation*}
	\dot V - \lambda^* \left\| {{\varepsilon _d}} \right\|^2 - \lambda^*_s \left\| {{\varepsilon _d}} \right\| \le -\frac{\mu}{2} \left\|x - x^*\right\|^2.
\end{equation*}
Similar to the proof in \cite{Ganghui2025arXiv}, it can be proved that $V$ is bounded, and therefore, $x$, $\hat{x}_i$, $\gamma_i$, and $\gamma_{is}$ are all bounded. According to \eqref{dnesa} and \eqref{relaxassum}, $\dot{x}$ is also bounded, which guarantees that $x$ is uniformly continuous. Then it follows from the Barbalat's Lemma \cite{H.K.Khalil2002} that $\mathop {\lim }\nolimits_{t \to \infty } \left\| x(t) - x^* \right\| = 0$.

\bibliographystyle{elsarticle-num}
\bibliography{autosam}           


\end{document}